\documentclass[twocolumn]{article} 

\usepackage[utf8]{inputenc} 
\usepackage{amsmath} 
\usepackage{graphicx} 
\usepackage{lipsum} 
\usepackage{geometry}
\usepackage{abstract} 
\usepackage{titlesec} 

\usepackage{authblk}
\usepackage{placeins}
\usepackage{float}
\usepackage{subfigure}
\usepackage{amssymb}
\usepackage[table,xcdraw]{xcolor}
\usepackage{makecell}
\definecolor{light-gray}{gray}{0.8}
\definecolor{ultramarine}{RGB}{71,108,154}
\usepackage[square,numbers]{natbib}
\usepackage{hyperref}
\usepackage{siunitx}
\usepackage{xcolor}
\usepackage{csquotes}
\usepackage{nicefrac}
\hypersetup{
    colorlinks=true,
    linkcolor=blue,
    filecolor=magenta,      
    urlcolor=blue,
    citecolor=blue,
}

\title{\Large 
\textbf{Freestanding perovskite and infinite-layer nickelate membranes}}

\author[1,\thanks{hoshang.sahib@ipcms.unistra.fr}]{Hoshang Sahib
}
\author[1]{Laurent Schlur}
\author[2]{Kumara Cordero}
\author[1]{Nathalie Viart}
\author[2]{David Pesquera}
\author[1,\thanks{daniele.preziosi@ipcms.unistra.fr}]{Daniele Preziosi}

\affil[1]{\small Université de Strasbourg, CNRS, IPCMS UMR 7504, F-67034 Strasbourg, France}
\affil[2]{\small Catalan Institute of Nanoscience and Nanotechnology, ICN2 Campus UAB 08193 Bellaterra, Barcelona, Spain}

\date{(Dated: \today)} 

\begin{document}

\twocolumn[
\maketitle
\vspace{-0.4cm}
\begin{abstract}
Following the discovery of superconductivity in hole-doped NdNiO$_2$ infinite-layer thin films, extensive research has been conducted particularly to compare these materials with cuprates. 
Superconductivity has also been observed in nickelate thin films with other rare-earth elements like Pr and La, but not in their bulk forms, suggesting a critical role for substrate-induced strain/interface or dimensionality effects. 
In this study, we use water-soluble (Ca,Sr)$_3$Al$_2$O$_6$ sacrificial layers to fabricate freestanding perovskite nickelate membranes and explore topotactic reduction without the SrTiO$_3$ substrate as a template. 
NdNiO$_3$-based heterostructure membranes transferred from a LaAlO$_3$ substrate exhibit better metallic behavior and higher hysteresis than those transferred from SrTiO$_3$, owing to the different strain states that the NdNiO$_3$ layer experience when grown on these two substrates. 
Despite the expected X-ray diffraction shifts, membranes reduced with CaH$_2$ display insulating characteristics, similar to bulk infinite-layer nickelates. 
Our findings strongly indicate that a template is necessary to stabilize a coherent and robust infinite-layer phase with optimal transport properties.
\end{abstract}
\vspace{0.1cm} 
\vspace{0.3cm} 
]
\saythanks 

\section*{Introduction}
Perovskite rare-earth nickelate heterostructures, $R$NiO$_3$ ($R$ being a rare-earth), have been subject of intense theoretical and experimental research efforts.
In particular a theoretical proposal assumed that those materials could mimic the high-Tc superconductivity of cuprates\cite{Anisimov1999}, for which a $Cu$-3d$e_g$ orbital selective Cooper pairing is believed to be responsible for the zero-resistance state\cite{Oles2019}.
Recently, soft-chemistry processes utilizing CaH$_2$ powder as a reducing agent have allowed the stabilization of a superconducting infinite-layer (IL) phase in hole-doped $R$NiO$_2$ thin films (with $R$ = Nd, Pr, and La) grown onto SrTiO$_3$ (STO) substrates\cite{li2019superconductivity,osada2020superconducting, zeng2022superconductivity}, which resemble cuprates by sharing both planar symmetry and a 3d$^9$ electron count.
The absence of bulk superconductivity, even under pressure\cite{Li2020Absence}, highlights the crucial role of substrate-induced strain/interface and/or dimensionality effects in stabilizing the superconducting phase.
To date, the nickelate IL phase is stabilized by optimizing the lattice mismatch between the perovskite and IL phases and the substrate. 
Specifically, the tetragonal IL phase ($a = 0.392\,\text{nm}, c = 0.328\,\text{nm}$ as measured for the undoped bulk compound\cite{HAYWARD-NNO}) is stabilized on STO ($a = 0.391\,\text{nm}$), to minimize the lattice mismatch (approximately -0.4\%). 
However, the STO imposes a large tensile strain on the perovskite precursor phase with a pseudocubic lattice parameter of $a_{pc} = 0.381\,\text{nm}$ (approximately +2.5\%). 
This large tensile strain leads to a larger content of oxygen vacancies if compared to a relatively small compressive strain (approximately -0.5\%) for growth onto LaAlO$_3$ (LAO) single crystal \cite{guo2020tunable}. In both cases a proper optimization process of the growth conditions is necessary to achieve bulk-like transport properties\cite{preziosi2017reproducibility}.

Moreover, stabilizing the Nd$_{1-x}$Sr$_{x}$NiO$_3$ perovskite phase for $x \geq 0.2$ onto STO is challenging due to Sr-segregation issues and an unfavorable valence state of the Ni ions\cite{krieger2022synthesis, Lee2020}. 
These factors necessitate a delicate balance of growth parameters to mitigate epitaxial mismatches for both the perovskite precursors and IL phases. 
A novel approach to overcome these challenges involves the development of freestanding nickelate membranes, which are free from substrate-induced strain, using a chemical-based etching method. 
This method includes growing the material onto a sacrificial layer, which is later removed via selective etching, allowing for the fabrication of oxide membranes while maintaining their structural integrity\cite{Pesquera2022}. 
Additionally, utilizing water-soluble sacrificial layers (Ca,Sr,Ba)$_3$Al$_2$O$_6$, which can be dissolved at room temperature without the use of any chemical pollutants, provides a versatile method to fabricate nickelate membranes. 
The lattice parameter of these layers can be adjusted from 0.382\,\text{nm} (100\% Ca) to 0.412\,\text{nm} (100\% Ba), thereby enhancing membrane quality and reducing crack formation\cite{Chiabrera2022, prodjosantoso2000synthesis}.

Here, we report on the fabrication of freestanding perovskite NdNiO$_3$ membranes using (Sr,Ca)$_3$Al$_2$O$_6$ sacrificial layers, and on the subsequent topotactic reduction after transferring onto silicon wafers to obtain IL NdNiO$_2$ membranes with good structural integrity. 
Although superconductivity is typically associated to the successful topotactic reduction of Sr-doped NdNiO$_3$ thin films, our focus on undoped NdNiO$_3$ is aimed at demonstrating the feasibility of stabilizing the IL phase after release.
Our experimental approach allow membranes extending up to 5x5 mm$^2$ to be seamlessly transferred onto various substrates, yielding crack-free flakes of millimeter size. 
X-ray diffraction measurements have demonstrated that the membranes retain the crystalline quality of the original seed films. 
Furthermore, we demonstrate the two-fold potential of using a sacrificial layer to stabilize the IL phase via topotactic reduction of nickelate membranes: 
(1) we could perform the growth of the precursor nickelate onto substrates with an ideal lattice mismatch ($i.e.$, LAO), therefore limiting the formation of extended defects triggered by the strain accommodation highly relevant in the case of STO-grown samples; 
(2) we could obtain first and important information about the substrate's influence on the IL stabilization, making a direct link with bulk preparation of IL nickelates.
\section*{Results and Discussion}
\begin{figure*}[ht!]
    \centering 
    \includegraphics[width=0.28\linewidth]{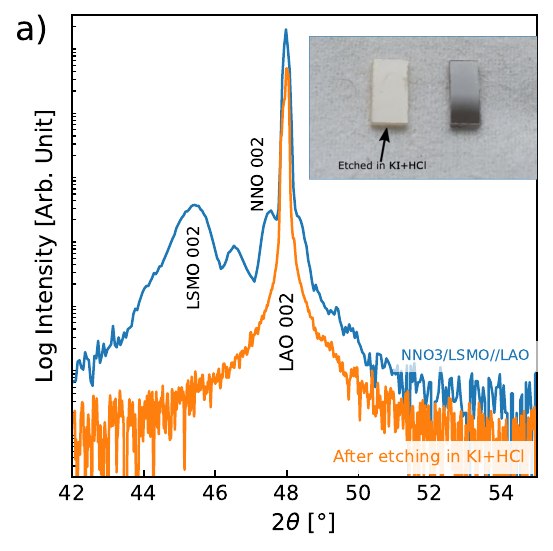}
    \includegraphics[width=0.3\linewidth]{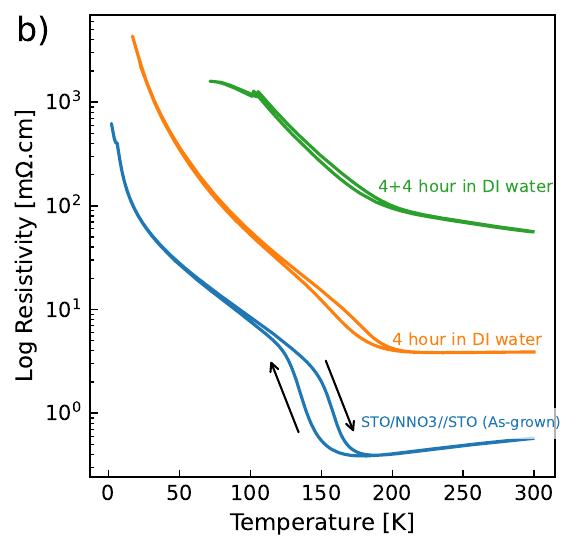}
    \includegraphics[width=0.29\linewidth]{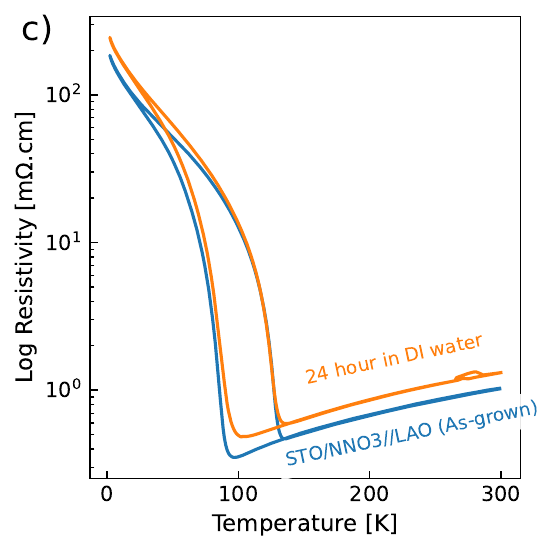}
    \caption{a) $2\theta-\theta$ scan of the 10 nm NNO3 layer on LSMO on LAO substrate before and after etching in a solution of 4 mg KI, 5 ml HCl (37\%) + 200 ml, with an inset showing a photograph of a 10 nm NNO3 layer on an STO substrate before and after etching. 
    b-c) Transport measurements of STO-capped NNO3 grown on STO and LAO substrates, respectively, before and after exposure to DI water.}
    \label{fig:In-KiHcl-water}
\end{figure*}
An important prerequisite for realizing freestanding crystalline membranes is the choice of the sacrificial layer, which must ensure a good lattice mismatch with both the target membrane layers and the substrate, while allowing for perfect etching selectivity.
As an example, here, we report on our experience using La$_{0.825}$Sr$_{0.175}$MnO$_3$ (LSMO) as a potential sacrificial layer to fabricate NdNiO$_3$ (NNO3) membranes. 
Motivated by the fact that LAO single crystals are ideal substrates for the growth of nickelate thin films due to the minimal strain \cite{Catalano2018}, we first optimized NNO3/LSMO//LAO heterostructures. 
Despite their high structural quality, we were unable to fabricate NNO3 membranes because the LSMO etchant (HCl+KI-based solution) was not selective, etching NNO3 similarly to LSMO. 
Figure~\ref{fig:In-KiHcl-water}a shows $\theta-2\theta$ X-ray diffraction (XRD) patterns for the NNO3/LSMO//LAO heterostructure before and after approximately two hours of etching. 
The XRD pattern, characterized by Laue oscillations around the LAO (00l) peaks due to the perfect NNO3-LAO lattice mismatch, completely disappears after etching. 
This confirms the non-selectivity of the HCl+KI-based solution, which was also tested with a bare NNO3//STO sample. 
The optical microscope images in the inset of Figure~\ref{fig:In-KiHcl-water}a shows NNO3//STO before and after etching, in which the NNO3 layer is completely removed.
Similar results were obtained (not shown) even when the NNO3 layers were encapsulated between two 10\,nm thick STO layers as a potential protection from direct contact with the LSMO etching solution.


To cope with this problem, we resorted to water-soluble sacrificial buffer layers already used in the literature to fabricate perovskite membranes \cite{Lu2016}. 
These are particularly promising for the fabrication of nickelate membranes due to their good structural compatibility with nickelates. 
For example, Ca$_3$Al$_2$O$_6$ (CAO) has a cubic unit cell (space group Pa$\bar{3}$) with a lattice parameter of 1.526\,nm, which closely matches the pseudocubic unit cells of both LAO and NNO3 (a$_{\text{CAO}}$ = 1.526  $\sim$ 0.381$\times$4\,nm).
Moreover, substituting Ca with Sr in specific ratios allows for tuning the CAO lattice parameter from $a$/4 $\sim$ 0.381\,nm for Ca$_3$Al$_2$O$_6$ to $a$/4 $\sim$ 0.396\,nm for Sr$_3$Al$_2$O$_6$ (SAO),
which dissolves in deionized (DI) water much faster (in a few minutes compared to approximately two days for CAO) due to the predominantly ionic Sr–O bonds \cite{prodjosantoso2000synthesis}. 
This extended dissolution time necessitates studying the effect of DI-water on the functional properties of NNO3, as protonation effects have been shown to play a role on the related transport properties \cite{Chen2022}. 
Figure~\ref{fig:In-KiHcl-water}b shows the temperature dependence of the resistivity measurements for STO-capped NNO3 thin films before and after immersion in DI-water of various duration. 
It is seen that properly optimized NNO3 thin films grown onto STO are very sensitive to DI-water, and the usual hysteretic metal-to-insulator transition (MIT) with an onset temperature of approximately 140\,K is significantly altered until the sample becomes fully insulating after eight hours of immersion. 
This behavior is absent when NNO3 is grown onto LAO substrates, where similar transport properties are observed even after one day in DI-water, as shown in Figures~\ref{fig:In-KiHcl-water}c. 
This compelling difference is most likely originating from the different strain state that the NNO3 thin films experience when grown onto STO (tensile) and LAO (slightly compressive) substrates which is found to affect the intrinsic content of oxygen vacancies \cite{guo2020tunable}.
To counteract the effect of DI-water on the fabricated membranes, we performed an oxygen post-annealing process as will be introduced later.

\subsection*{Epitaxial growth of the heterostructures}

\begin{figure}[hb!]
    \centering 
    \includegraphics[width=0.62\linewidth]{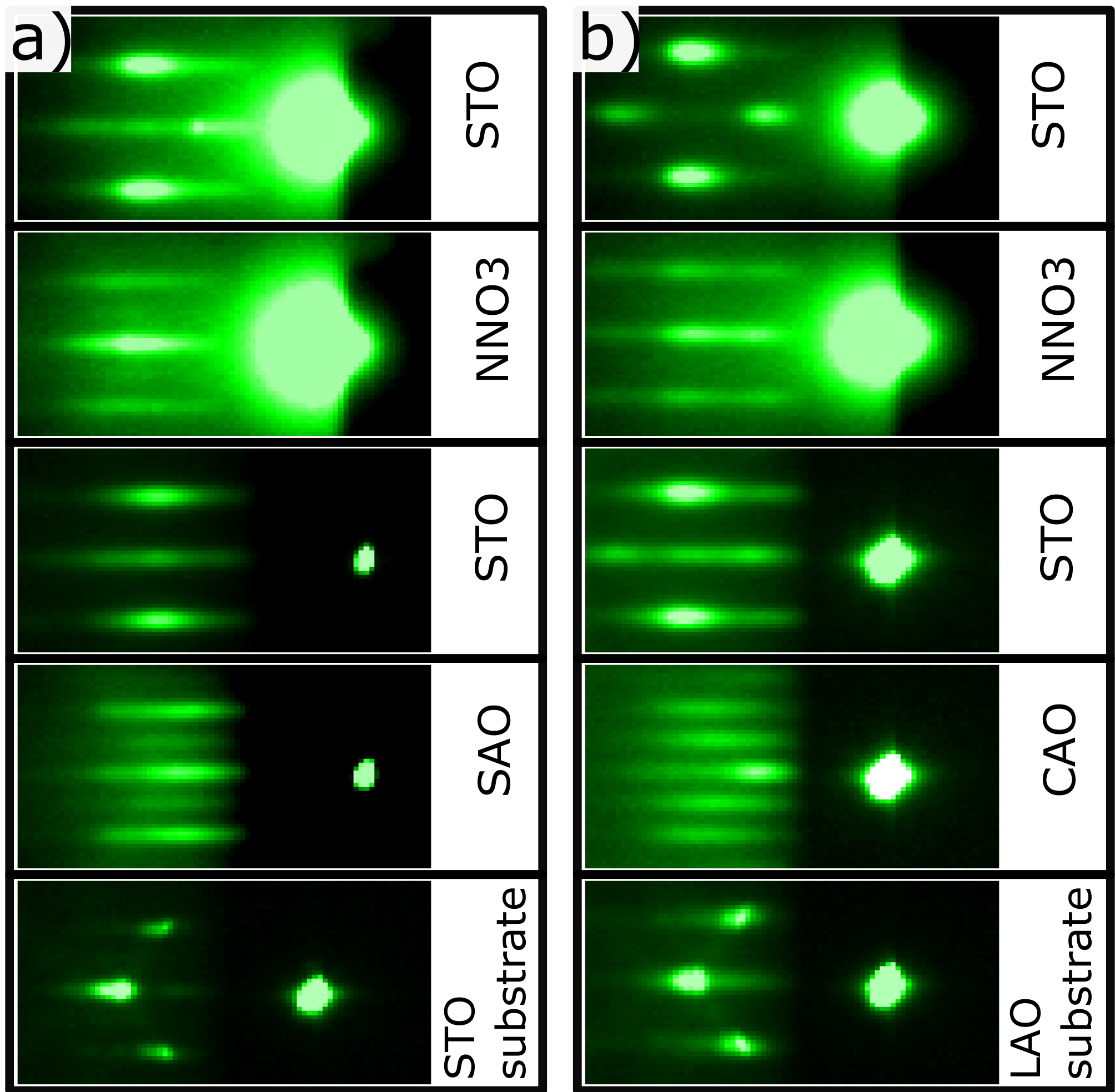}
    \includegraphics[width=0.34\linewidth]{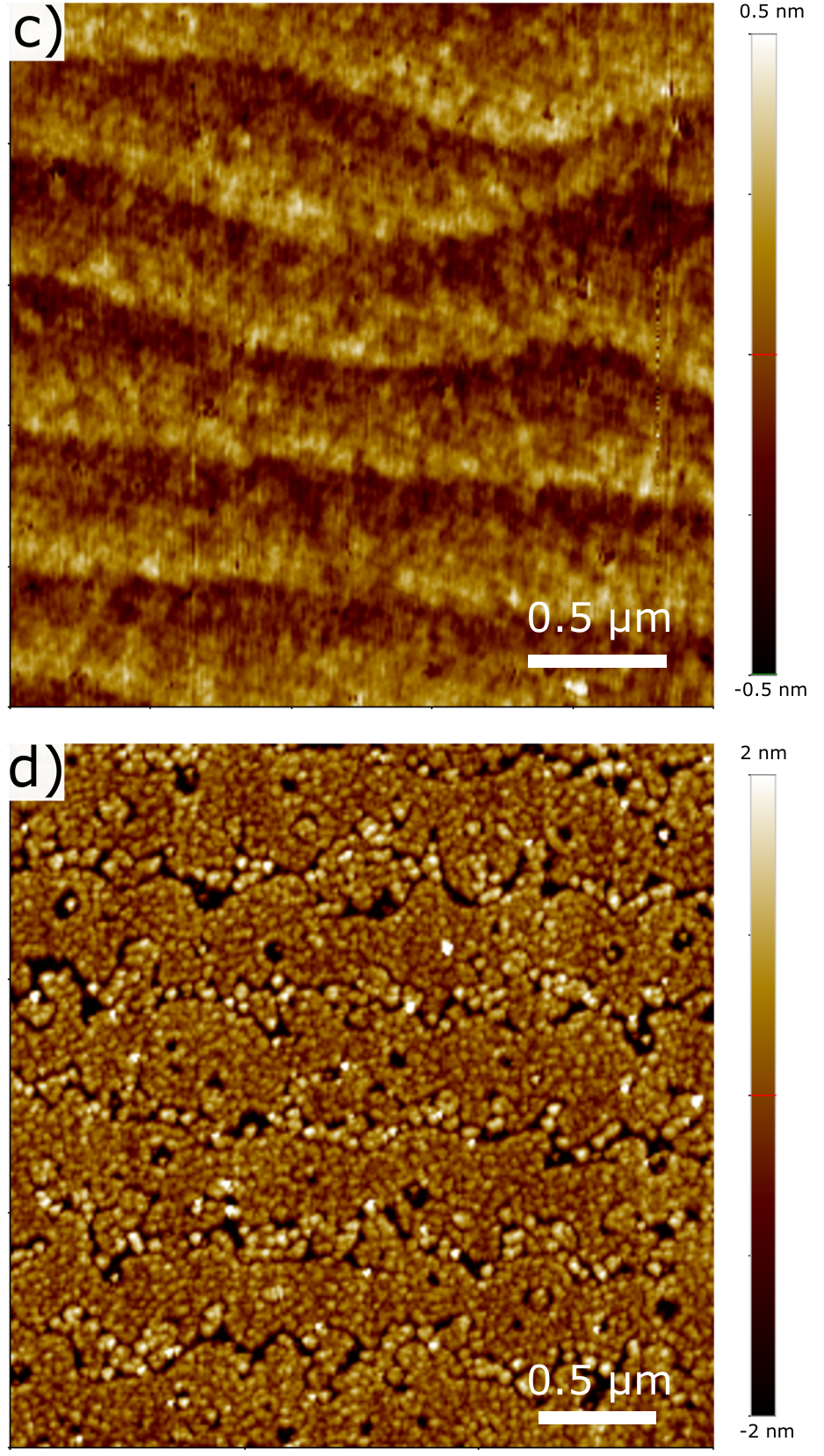}
    \caption{a--b) RHEED diffraction patterns for CAO-LAO and SAO-STO HTs along the (001) direction of the substrates upon the completion of each layer's growth. 
    The order of the RHEED images is from bottom to top, following the structure of the samples at each corresponding stage. 
    c--d) AFM topography images for CAO-LAO and SAO-STO HTs acquired after the growth, respectively.
    }
    \label{fig:Rheed-afm}
\end{figure}
We used a pulsed laser deposition technique to grow NNO3-based heterostructures (HTs) using several sacrificial layers, such as $\mathrm{Sr_3Al_2O_6}$ (SAO), $\mathrm{Ca_3Al_2O_6}$ (CAO), and $\mathrm{Ca_2Sr_1Al_2O_6}$ (CSAO) on both STO and LAO substrates (please refer to the Experimental Section for details). 
The NNO3 films directly grown onto the sacrificial layers appeared to exhibit modified transport properties, and the sacrificial layers became insoluble in water, suggesting that interdiffusion modifies the electronic and chemical properties of both layers. Therefore, the 10\,nm thick NNO3 films were encapsulated between two layers of STO, approximately 10--25 unit cells thick, to prevent potential interdiffusion of ions between the sacrificial layers and NNO3 \cite{baek2017mapping}.
Reflection high-energy electron diffraction (RHEED) patterns taken after the growth of sacrificial, STO bottom, and NNO3 layers showed 2D streaky features, indicating clear layer-by-layer growth and an atomically flat surface, as illustrated in Figures~\ref{fig:Rheed-afm}a--b for the STO/NNO3/STO/SAO//STO and STO/NNO3/STO/CAO//LAO HTs, hereafter referred to as SAO-STO and CAO-LAO, respectively. Results obtained using the CSAO sacrificial layer are shown in Figures~\ref{fig:SI-RheedAfm}a--b.
After the growth of the top STO layer, the RHEED pattern exhibited a mix of 2D streaks and 3D transmission spots, indicating that the surface had developed a partial 3D morphology. 
This was also confirmed by Atomic Force Microscopy (AFM) measurements, with root-mean-square (rms) surface roughness ranging from 0.2\,nm to 0.7\,nm, as shown in Figures~\ref{fig:Rheed-afm}c--d.

\begin{figure*}[ht!]
    \centering 
    \includegraphics[width=0.42\linewidth]{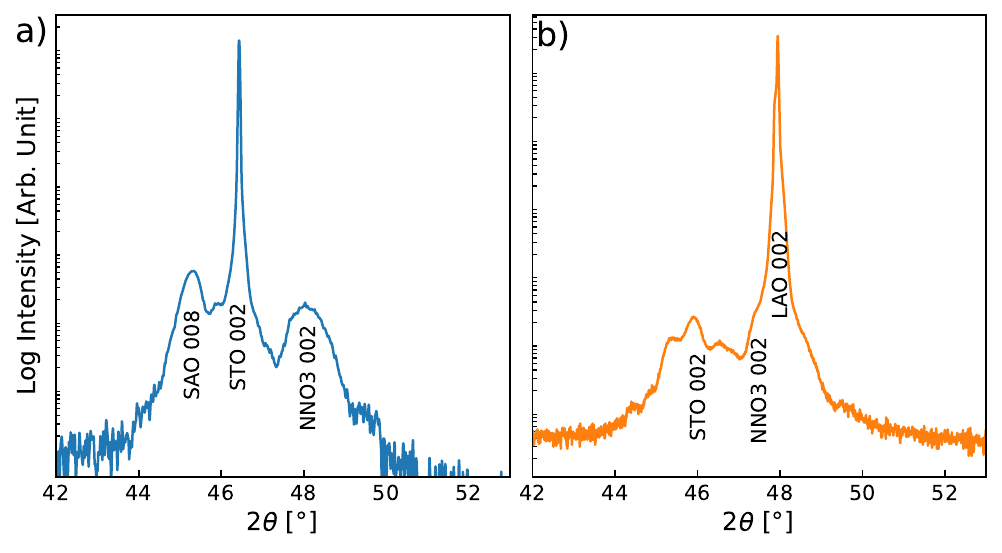}
    \includegraphics[width=0.33\linewidth]{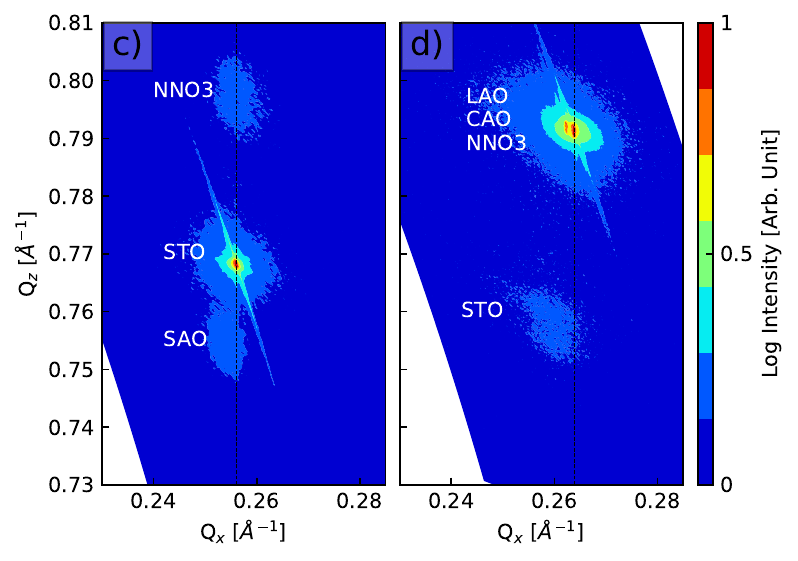}
    \includegraphics[width=0.215\linewidth]{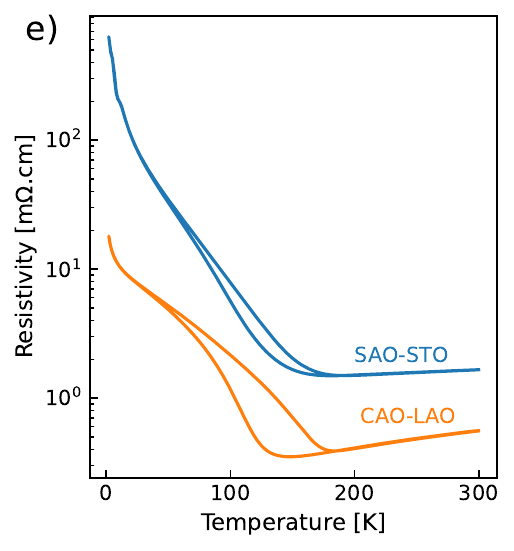}
    \caption{Structural characterization of the as-grown heterostructures. $\theta-2\theta$ scans of the as-grown a) SAO-STO and b) CAO-LAO heterostructures. 
    Reciprocal space maps around the STO and LAO -103 asymmetric reflections for the c) SAO-STO and d) CAO-LAO heterostructures.
    e) Transport measurements of both SAO-STO and CAO-LAO heterostructures.
}
    \label{fig:th2th-rsm-epitaxial}
\end{figure*}

Figures~\ref{fig:th2th-rsm-epitaxial}a--b show the $\theta$--$2\theta$ XRD symmetric scans of both the CAO-LAO and SAO-STO HTs. 
The clear diffraction peaks of the sacrificial layers and the STO/NNO3/STO HTs confirm that all layers are grown with good crystalline quality. 
The STO multipeak structure for the CAO-LAO HTs might be associated with a partial strain gradient, which was also observed for STO/NNO3/STO/CSAO//LAO (CSAO-LAO) HTs (see Figure~\ref{fig:SI-th2th-rsm-epitaxial}a). 
Reciprocal space mapping presented in Figures~\ref{fig:th2th-rsm-epitaxial}c-d, acquired around the asymmetric -103 reflection of the substrates, show that the SAO sacrificial layer is not fully in-plane strained to the STO; surprisingly, the subsequent layer of NNO3 exhibits the same in-plane lattice parameter as the STO substrate. 
For the CAO-LAO HTs, both the sacrificial layer and NNO3 are fully strained to the LAO, as expected from the good lattice mismatches, while the STO is partially relaxed. 
To attest to the quality of the NNO3 films, we also performed temperature-dependent transport measurements, and as reported in Figure~\ref{fig:th2th-rsm-epitaxial}e, both HTs exhibit the characteristic features of perovskite nickelate thin films with metallic behavior at high temperatures followed by a hysteretic metal-to-insulator transition below approximately 150\,K, with a resistance change of at least two orders of magnitude \cite{torrance1992systematic}. 
In particular, the CAO-LAO HT shows better metallic behavior, with an overall resistivity at 300\,K lower than 1\,m$\Omega$$\cdot$cm, and a larger hysteresis, likely resulting from the better lattice match with the LAO single crystal and the reduced internal strain state of the HTs, which leads to fewer extended defects and/or oxygen vacancies. 
Overall, these results indicate that the sacrificial layers with the adjacently grown STO thin films do not affect the quality of the NNO3 thin films, which exhibit good epitaxial growth. 
It is worth mentioning that similar transport properties for NNO3 are measured for CSAO-LAO HTs, while STO/NNO3/STO/SAO//LAO (SAO-LAO) HTs resulted in a completely insulating behavior of the NNO3, even though the RHEED patterns indicated 2D-like growth of the layers (see Figure~\ref{fig:SI-th2th-rsm-epitaxial}c).

\subsection*{Epitaxial lift-off (membrane release)}
\begin{figure}[hb!]
    \centering
    \includegraphics[width=0.99\linewidth]{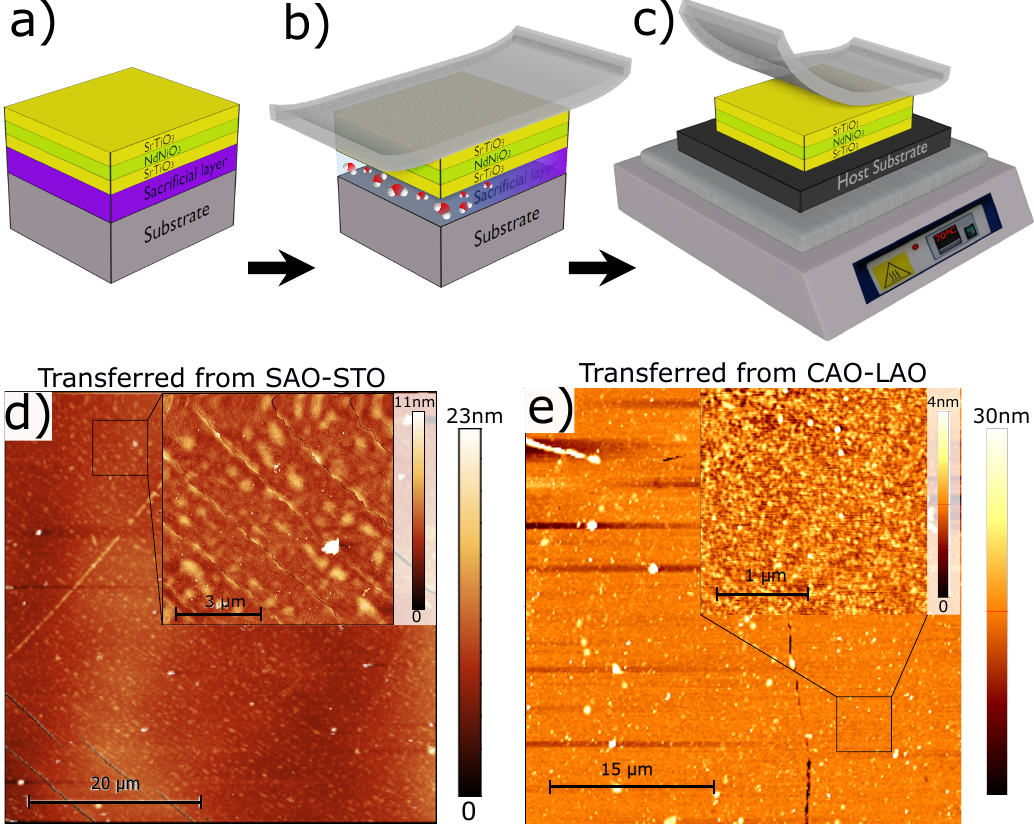}
    \caption{a--c) Freestanding membranes fabrication process flow. 
    a) Epitaxial growth of the sacrificial (water-soluble) layer and the STO/NNO3/STO heterostructure. 
    b) Fixing the polymer support on sample surface and etching in de-ionized water. 
    c) Transfer to the host substrate and peel-off of the polymer support. 
    d-e) AFM topography images of the heterostructure transferred from SAO-STO and CAO-LAO respectively.}
    \label{fig:fabrication_process_flow}
\end{figure}
Figure~\ref{fig:fabrication_process_flow}a--c sketch the process step-flow for fabricating freestanding membranes. 
First, water-soluble sacrificial layers are grown onto STO and/or LAO single crystals, followed by the growth of the STO/NNO3/STO heterostructure. 
A polymer support (PDMS), attached to a glass slide, is cut into pieces larger than the sample surface, and brought into conformal contact with the sample surface. 
Then, the stack of polymer and sample is immersed in DI-water to selectively etch the sacrificial layer. 
The etching time for the sacrificial layers varies from one to two hours for SAO, up to two days for CSAO and CAO. 
After verifying the complete etching of the sacrificial layer, the glass+PDMS+heterostructure stack is released from the substrate, then placed in contact with the recipient substrate (SiO$_2$/Si) using a transfer system with micrometer XYZ stages. 
The SiO$_2$/Si substrate had been previously ultrasonically cleaned in acetone and isopropyl alcohol. 
After annealing the entire stack in air at 70$^\circ$C for 10 minutes to promote adhesion on the SiO$_2$/Si substrate, the PDMS stamp is then carefully peeled off by raising the glass+PDMS support using one of the micrometer stages of our system.

\begin{figure*}[ht!]
    \centering
    \includegraphics[width=0.28\linewidth]{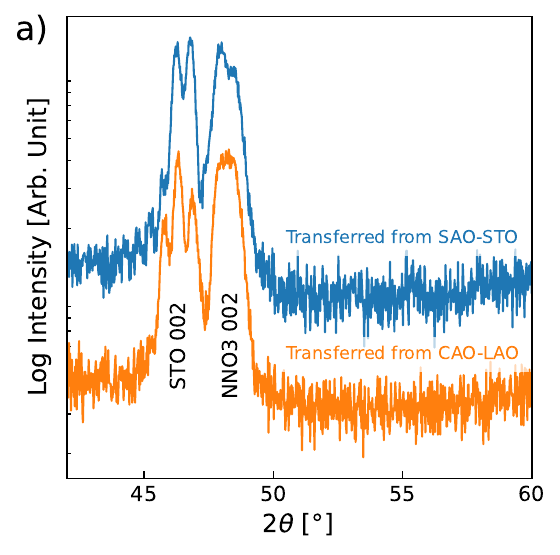}
    \includegraphics[width=0.39\linewidth]{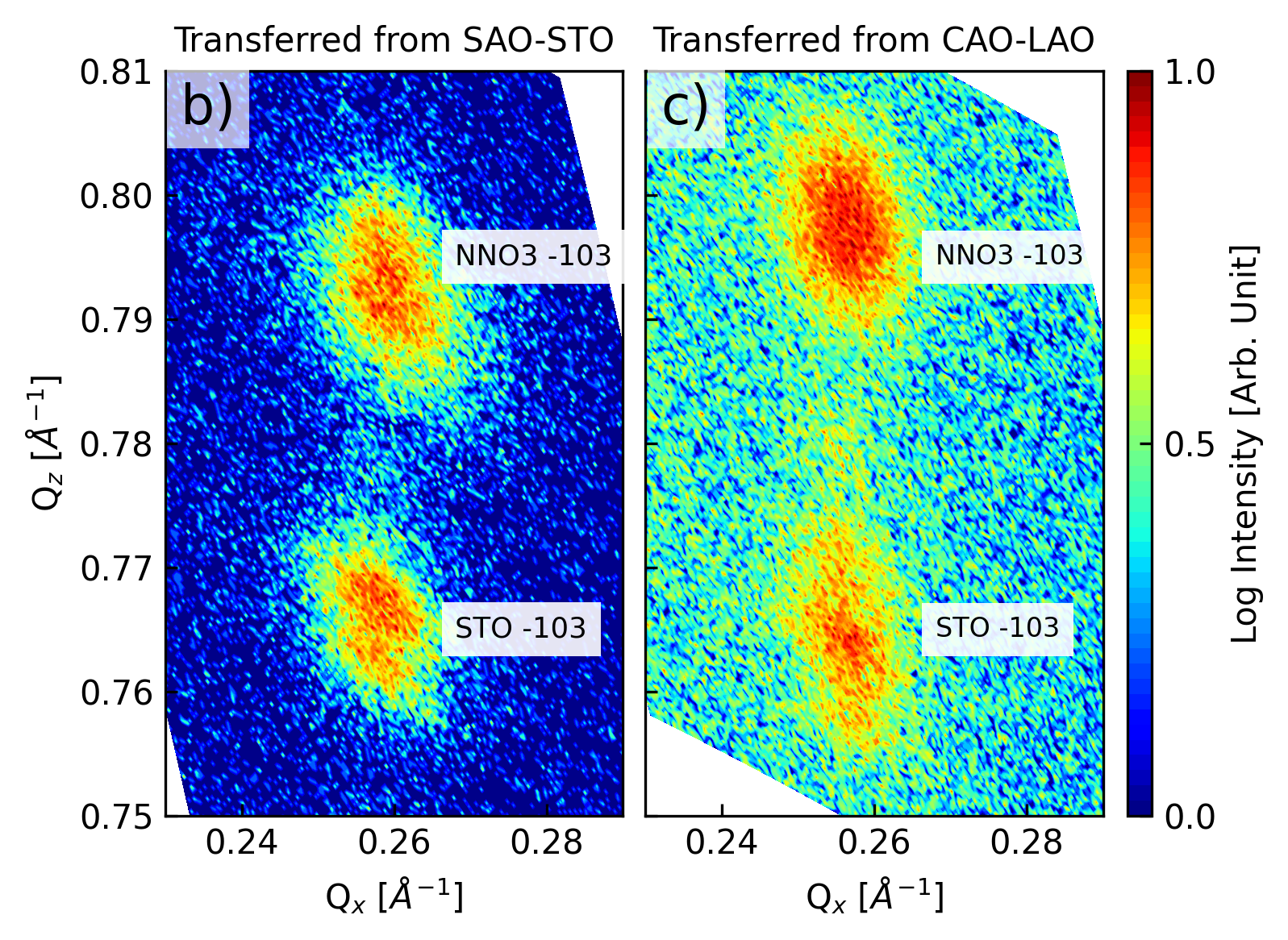}
    \includegraphics[width=0.3\linewidth]{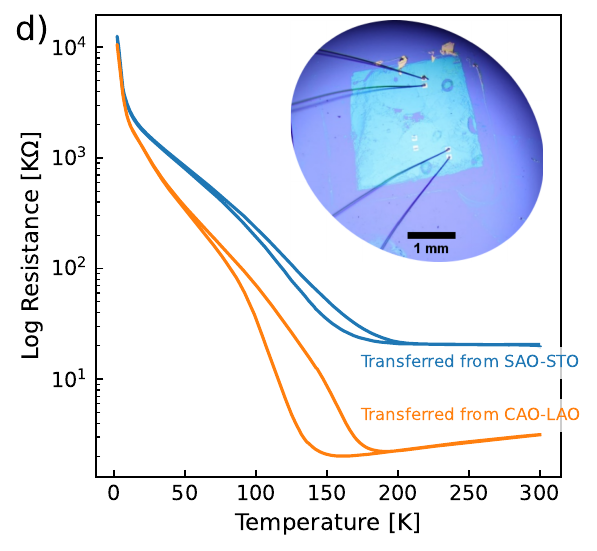}
    \caption{a) $2\theta-\theta$ scan and b--c) Reciprocal space maps around NNO3 and STO -103 reflection of the STO/NNO3/STO heterostructures transferred from SAO-STO and CAO-LAO, respectively. d) Transport measurements of STO/NNO3/STO heterostructures membranes after transferring on to SiO2/Si wafer and annealing for 3 hours at 300$^\circ$C under oxygen flow. Insets: optical image of the freestanding membrane on SiO2/Si substrate with micro Au-electrodes and Al wire bonds.}
    \label{fig:th2thAndRsm-exfoliated}
\end{figure*}
Depending on the lattice parameters of the sacrificial layer and substrate, the resultant membranes exhibited different characteristics. 
The membranes obtained by dissolving SAO and CAO sacrificial layers were characterized by a complete transfer onto the polymer support, yet with notably different numbers and sizes of cracks. 
Membranes released from SAO-STO exhibited parallel micrometer-scale cracks in one direction, which were only visible by AFM, while the CAO-LAO membranes displayed fewer cracks and lacked any preferential direction, as shown in Figure~\ref{fig:fabrication_process_flow}d-e. 
Such differences in crack formation between these two types of membranes can be attributed to the strain state and number of defects; a low lattice mismatch between NNO3, CAO, and the LAO substrate permits coherent epitaxial growth of the films, which was confirmed by RSM measurements (see Figure~\ref{fig:th2th-rsm-epitaxial}c-d). 
On the other hand, membranes released from CSAO-LAO tended to bend upon themselves and curled into microscopic rolls, as shown in the SEM image of Figure~\ref{fig:SI-th2thAndRsm-exfoliated}c.

To further study the structural integrity of the freestanding membranes, we conducted XRD $\theta-2\theta$ symmetric and asymmetric scans of the membranes transferred onto SiO$_2$(300 nm)/Si wafers. 
After transfer, the membranes were annealed for 3 hours at 300$^\circ$C under an oxygen flow to mitigate possible protonation effects. 
In Figure~\ref{fig:th2thAndRsm-exfoliated}a, we present the $\theta-2\theta$ scan of STO/NNO3/STO heterostructure membranes transferred from SAO-STO and CAO-LAO HTs. 
It is seen that all membranes exhibit (002) diffraction peaks for STO and NNO3, similar to those of the as-grown HTs before release, confirming the conservation of crystalline integrity for both NNO3 and STO layers. 
The three side-by-side peaks for the STO layers may result from the  Laue oscillations or superposition of the diffraction peaks of the bottom and upper STO layers, as already shown for the XRD measurements of the HTs before dissolving them in DI-water.
The RSM measurements performed on the membranes showed that, after release, the three layers of the STO/NNO3/STO HTs remain epitaxially coupled, resulting in an equilibrium strain state, as shown in Figure~\ref{fig:th2thAndRsm-exfoliated}b-c. 
The final strain of the oxide layer in the membranes results from a trade-off between their elastic properties, and the initial strain withstanding during growth of the oxide stack and assured by the choice of the substrate  \cite{pesquera2020beyond}. 
As a result of this process, in all membranes, the NNO3 layer is strained to the STO buffer layers, and conversely the STO layer is finely strained to NNO3. Interestingly, the NNO3 thin film grown onto LAO do not show the expected bulk values, but during release the in-plane lattice parameter is partially adjusted to the STO one. The lattice parameters and the calculated in-plane and out-of-plane strains in the membranes extracted from the RSM data are listed in Table~\ref{tab:param}.

In Figure~\ref{fig:th2thAndRsm-exfoliated}d, we present the electrical transport properties of the membranes, which were measured after the deposition of gold electrodes and aluminum wire bonding (see inset). 
The membranes preserved their transport properties that were shown in Figure~\ref{fig:th2th-rsm-epitaxial}e, with both the high-temperature metallic behavior and the low-temperature insulating phase, featuring a well-defined metal-to-insulator transition of the NNO3 layer and similar hysteresis as already reported for the grown HTs.
These results are highly reproducible, as can be seen in the Figure~\ref{fig:SI-multiple-samples} in supporting information.
\begin{table}[!ht]
    \footnotesize
    \caption{Lattice parameters and strain states of the STO/NNO3/STO freestanding membrane heterostructures transferred from SAO-STO to CAO-LAO. Note: The strain state in the membranes depends on the thickness of the STO buffer layers; here, the STO thickness is $\sim$ 10 nm.}
        \centering
        \begin{tabular}{c|c|c|c|c}
\hline 
\rowcolor[HTML]{EFEFEF}
\makecell{Material} & \makecell{a-axis\\$\left[Å\right]\pm0.01$} & \makecell{c-axis\\ $\left[Å\right]\pm0.01$} & \makecell{ip strain\\ $\left[\%\right]$} & \makecell{oop stain\\ $\left[\%\right]$}
        \\ \hline
         \makecell{STO of \\SAO-STO} & 3.87 & 3.91 & -0.9 & 0.13  \\ \hline
		 \makecell{NNO of \\SAO-STO} & 3.87 & 3.78 & 1.63  & -0.74  \\ \hline
         \makecell{STO of \\CAO-LAO} & 3.89 & 3.92 & -0.38 & 0.51   \\ \hline
		 \makecell{NNO of \\CAO-LAO} & 3.89 & 3.76 & 2.15  & -1.26  \\ \hline
				 \end{tabular}
    \label{tab:param}
\end{table}

\subsection*{Topotactic reduction of the membranes}
\begin{figure}[t!]
    \centering 
    \includegraphics[width=0.625\linewidth]{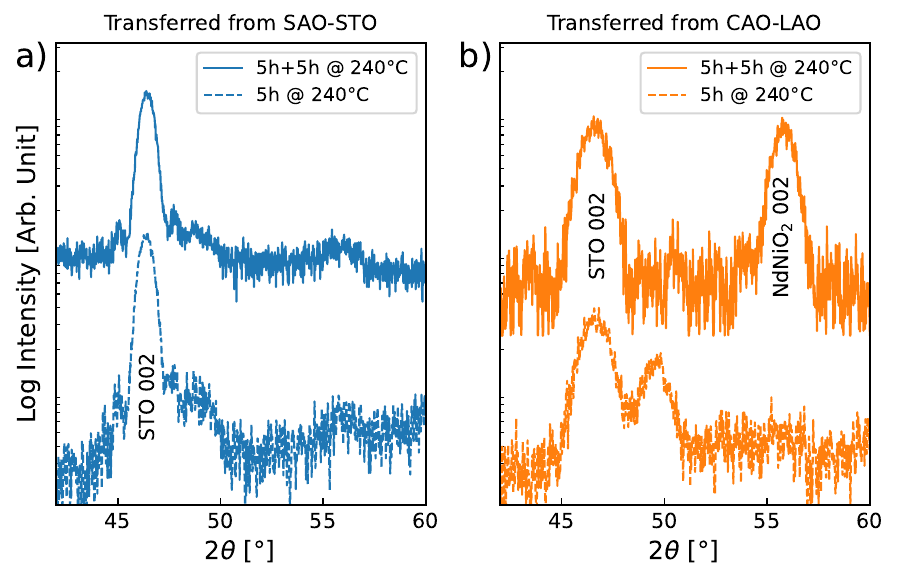}
    \includegraphics[width=0.36\linewidth]{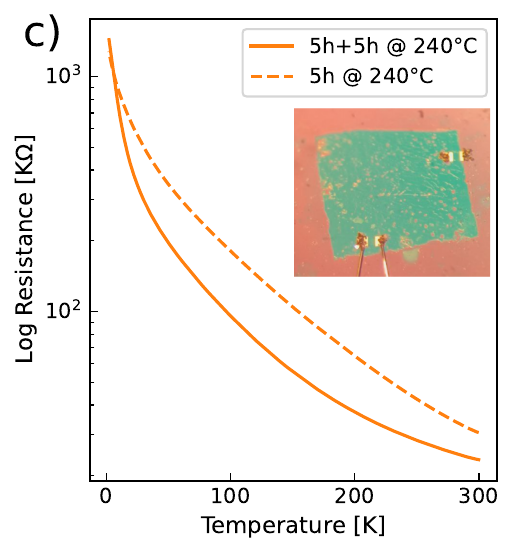}
    \caption{a--b) $2\theta-\theta$ scan of the STO/NNO3/STO heterostructure membranes after topochemical reduction for 5 hours at 240$^\circ$C, followed by an additional 5 hours of reduction at the same temperature, for membranes transferred from a) SAO-STO and b) CAO-LAO. 
    c) Transport measurement of the freestanding infinite-layer membranes transferred from CAO-LAO; inset: optical microscope image of a reduced membrane after bonding.
    }
    \label{fig:th2th-transport-exfoliated-topo}
\end{figure}

After the successful fabrication of the STO/NNO3/STO membranes, we performed a topochemical reduction by placing them in an evacuated silica tube sealed with a membrane-valve, where 0.5\,g of CaH$_2$ powder was put in direct contact with them, as used in prior studies\cite{krieger2022synthesis}. 
The process was optimised by a systematic series of steps, with $\theta-2\theta$ scans and electrical transport measurements performed ex situ after each step to assess the degree of topotactic reduction. 
A temperature of 240$^\circ$C was found to be optimal for reduction, high enough to enable the reduction and low enough to avoid any damage to the membranes.
In Figure~\ref{fig:th2th-transport-exfoliated-topo}, we present $\theta-2\theta$ scans and electrical transport measurements of two steps of the topotactic reduction process. 
After a first step of reduction, the NNO3(002) XRD diffraction peak considerably decreased in intensity for membranes obtained from SAO-STO HTs, while for CAO-LAO membranes, the XRD peak, still intense, began to shift towards higher $2\theta$ values as expected for a proper stabilization of the IL phase. 
An additional step of reduction resulted in a complete shift to the typical $2\theta$ value expected for IL films, for the membranes obtained from CAO-LAO HTs, with substantial intensity (Figure~\ref{fig:th2th-transport-exfoliated-topo}b). The c-axis lattice parameter of the membranes is found to be $\approx 0.330 \pm 0.001$\,nm, which is very similar to what we measure for unreleased NdNiO$_2$ thin films. 
Notably, the $\theta-2\theta$ scans, give the relevant information that the stabilization of the IL phase is obtained only in the case of membranes released from the CAO-LAO HTs while the ones obtained from the SAO-STO HTs practically decomposed, as demonstrated by the almost complete absence of the IL diffraction peak in Figure~\ref{fig:th2th-transport-exfoliated-topo}a. This disparity could be ascribed to tiny differences in the related perovskite precursor phases. In particular, as already highlighted above the membranes released from CAO-LAO HTs showed better metallic behavior and a larger hysteresis (Figure~\ref{fig:th2thAndRsm-exfoliated}e), while from the Table~\ref{tab:param}, we can note that only for CAO-LAO membranes the STO lattice parameter is closer to the bulk one, therefore exerting a lower compressive strain on the forming IL tetragonal crystal structure, if compared to the one obtained in the case of SAO-STO membranes.
Although the XRD peak indicated the formation of a high crystalline quality IL phase for the CAO-LAO membranes, which exhibited a full shift of the XRD peak, we consistently measured insulating transport properties.
This insulating behavior cannot be attributed to further cracks formation (inset of Figure~\ref{fig:th2th-transport-exfoliated-topo}c) during the topotactic reduction process, since 
the $\sim 20 \, \text{k}\Omega$ contact resistance is the one usually encountered in the case of IL thin films.
On the other side, from the structural point of view we observed a rather large decrease of the c-axis lattice parameter of the STO buffer layers after reduction. In particular, the STO c-axis modified from $0.392 \, \text{nm}$ to $0.3896 \, \text{nm}$, indirectly indicating an increase of the in-plane lattice parameter of the STO layer.
Therefore, the fully insulating behavior of IL membranes, which is similar to what has been measured for bulk IL \cite{Li2020Absence}, points to the experimental fact that a greater strain effect from the STO buffer layer (or substrate) is necessary to enable a more coherent stabilization of the IL phase with the expected transport properties.
\section*{Conclusion}
Our study confirm the use of water-soluble (Ca,Sr)$_3$Al$_2$O$_6$ as an appropriate epitaxial sacrificial layers for the growth of high-quality, single-crystal STO-encapsulated NdNiO$_3$ (NNO3) thin film heterostructures. 
By varying the composition of Ca and Sr, we successfully grew NNO3 thin films onto different substrates, showing that their structural and transport properties are similar to those observed for directly grown films. 
However, a buffer layer between the sacrificial layer and the perovskite thin films was necessary to limit problems of ion interdiffusion.
The dissolution of the sacrificial layers in DI-water allowed for the transfer of NNO3 heterostructure membranes from single-crystal substrates to any other support, such as SiO$_2$/Si. 
Encapsulating the NNO3 layer between STO layers and applying post-processing treatments, $i.e.$, oxygen annealing of the membranes, enabled the preservation of the intrinsic properties of the nickelate membranes post-release.
By means of topotactic reduction process via CaH$_2$ powder as the reducing reagent, 
IL membranes could be obtained only from CAO-LAO HTs 
which, nevertheless, displayed a fully insulating behaviour similarly to bulk IL nickelates. From our structural study it appeared clear that the STO buffer layers during the topotactic reduction re-adjusting their lattice parameters to the IL-phase one ($i.e.$ a = 0.392\,nm), could not provide the optimal (compressive) strain state to the IL-phase, which might be the necessary and sufficient condition to grant the ideal transport properties of IL nickelate thin films. 
Finally, our results highlight the significant effect of the substrate epitaxial strain on the structure and electronic properties of nickelate membranes, the formation of the IL phase, and the environmental stability on the performance of nickelate systems. 
They open possibilities to adjust the substrate-induced strain to optimize each different phase, NNO3 and NNO2, separately.
We are confident that the optimization of the sandwich material, either in its nature or thickness, will allow fine-tuning of the different strains and lead to freestanding conducting IL membranes.
\section*{Experimental Section:}
\paragraph{\textbf{Epitaxial film fabrication:}}
The epitaxial growth of NNO3-based  HTs onto STO and LAO substrates was done by pulsed laser deposition using a 248 nm KrF excimer laser, utilizing single-crystal targets from Toshima Manufacturing CO LTD. 
The STO substrates were prepared by etching in an NH$_4$F-buffered HF solution, then annealed for 2 hours at 950$^\circ$C in air to obtain a well-defined TiO$_2$-terminated step-terraced surface. 
In contrast, the LAO substrates were annealed for 10 hours at 1000$^\circ$C to achieve an AlO$_2$-terminated step-terraced surface. 
Prior to growth, each substrate was pre-annealed for 1 hour at 800$^\circ$C under a pressure of 0.3 mbar in an oxygen flow to ensure a sharp step-and-terrace surface. 
Approximately 20 nm $\mathrm{Sr_3Al_2O_6}$ (SAO), $\mathrm{Ca_3Al_2O_6}$ (CAO), and $\mathrm{Ca_2Sr_1Al_2O_6}$ (CSAO) sacrificial layers were grown on top of the STO(LAO) (001) substrates at a substrate temperature of 800$^\circ$C and an oxygen partial pressure of P$_{O_2}$ = $1.2 \times 10^{-5}$ mbar, using a laser fluence of 1.2 J/cm$^2$ and a $1 \times 1.4$ mm$^2$ laser spot size on the targets. 
The bottom STO buffer layers (6-10 nm) were grown on the sacrificial layers under the same growth conditions. 
Next, the 10 nm NNO3 layers were grown on top of the STO buffer layers at a temperature of 675$^\circ$C and P$_{O_2}$ = 0.3 mbar, with a laser fluence of 3.3 J/cm$^2$ and a $1 \times 1.4$ mm$^2$ laser spot size. 
Finally, the top STO layer was grown at a T of 575$^\circ$C and P$_{O_2}$ = 0.3 mbar, with a laser fluence of 1.2 J/cm$^2$ and a $1 \times 1.4$ mm$^2$ laser spot size. 

\paragraph{\textbf{Transfer of freestanding membranes:}}
A flexible polymer (PDMS) was adhered to the heterostructure surface (Figure~\ref{fig:fabrication_process_flow}b), and then the assembly was immersed in room-temperature de-ionized water to dissolve the sacrificial layers. 
After dissolution, the freestanding membranes were transferred to SiO$_2$/Si substrates by placing and pressing the membrane with PDMS onto the heated (70$^\circ$C) silicon wafer (Figure~\ref{fig:fabrication_process_flow}c). 
By peeling off the PDMS support, the membranes remain on the wafers and detach from the support.

\paragraph{\textbf{Characterization:}}
The surface morphology of the samples was acquired using a Park XE7 (Park System) atomic force microscope (AFM) in true non-contact mode. 
The XRD data were collected using Bruker D8 Advance and PANalytical X’Pert Pro diffractometers with a Cu-K$\alpha$ radiation source (1.54056 Å). 
The transport properties of the heterostructures on the single-crystal substrates were studied using a Dynacool system (Quantum Design) and measured using the van der Pauw method on 5$\times$5 mm$^2$ samples by applying current amplitudes of 10 $\mu$A. 
To measure the transport of the transferred membrane flakes, Au electrodes were first fabricated by photolithography; then, the electrodes were bonded, and the transport measurements were conducted in two-point contact mode.

---------------------------------------

\textit{Note: During the preparation of this manuscript, we became aware of two pre-prints on infinite-layer freestanding membranes of La$_{0.80}$Sr$_{0.20}$NiO$_2$ \cite{yan2024superconductivity} and Nd$_{1-x}$Sr$_x$NiO$_3$ \cite{lee2024millimeter}. 
}
\section*{Acknowledgments}

This work was funded by the French National Research Agency (ANR) through the ANR-JCJC FOXIES ANR-21-CE08-0021. This work was also done as part of the Interdisciplinary Thematic Institute QMat, ITI 2021 2028 program of the University of Strasbourg, CNRS and Inserm, and supported by IdEx Unistra (ANR 10 IDEX 0002), and by SFRI STRAT’US project (ANR 20 SFRI 0012) and EUR QMAT ANR-17-EURE-0024 under the framework of the French Investments for the Future Program. D. Pesquera and K.CORDERO acknowledge Grant No. PID2022-140589NB-I00 funded by  MCIN/AEI/10.13039/501100011033. D. Pesquera acknowledges funding from ‘la Caixa’ Foundation fellowship (ID 100010434). The ICN2 is funded by the CERCA programme / Generalitat de Catalunya and by the Severo Ochoa Centres of Excellence Programme, funded by the Spanish Research Agency (AEI, CEX2021-001214-S).
\bibliographystyle{unsrtnat}
\bibliography{refs.bib}
\onecolumn
\appendix

\section*{\fontsize{14pt}{18pt}\selectfont Supporting Information}

\renewcommand{\thefigure}{S\arabic{figure}}

\subsection*{\Large \textbf{Title: Freestanding perovskite and infinite-layer nickelate membranes}}

\noindent \textbf{Hoshang Sahib$^1$, Laurent Schlur$^1$, Kumara Cordero$^2$, Nathalie Viart$^1$, David Pesquera$^2$, and Daniele Preziosi$^1$,}

\vspace{0.2cm} 

$^1$ Université de Strasbourg, CNRS, IPCMS UMR 7504, F-67034 Strasbourg, France.

$^2$ Catalan Institute of Nanoscience and Nanotechnology, ICN2 Campus UAB 08193 Bellaterra, Barcelona, Spain

\vspace{0.3cm}  
\begin{figure}[ht!]
    \centering 
    \includegraphics[width=0.38\linewidth]{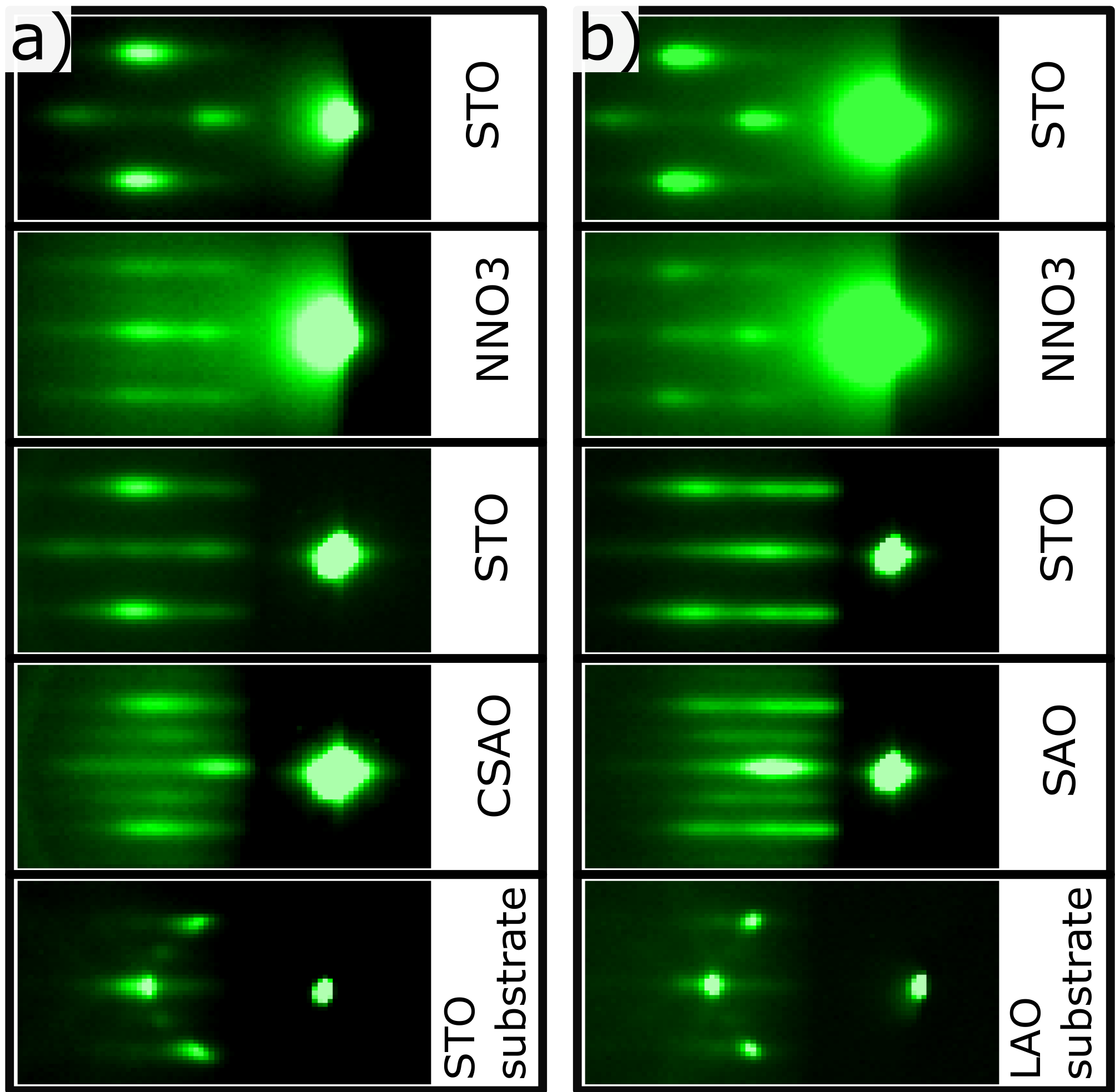}
    \includegraphics[width=0.21\linewidth]{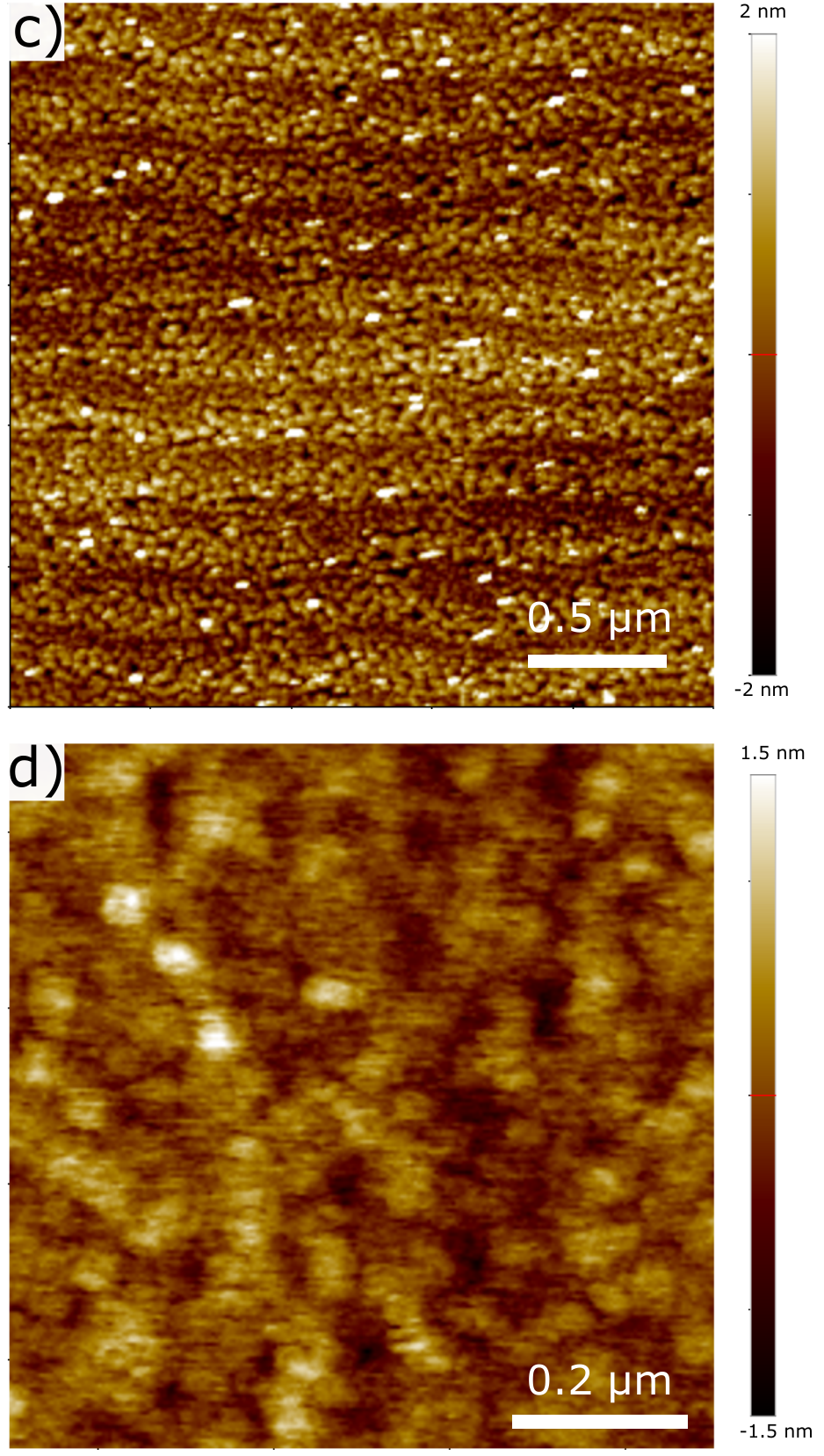}
    \caption{a--b) RHEED diffraction patterns for CSAO-LAO and SAO-LAO heterostructures  along the (001) direction of the substrates upon the completion of each layer's growth. 
    c-d) AFM topography images of CSAO-LAO and SAO-LAO HTs acquired after the growth, respectively.
    }
    \label{fig:SI-RheedAfm}
\end{figure}
\begin{figure}[ht!]
    \centering 
    \includegraphics[width=0.3\linewidth]{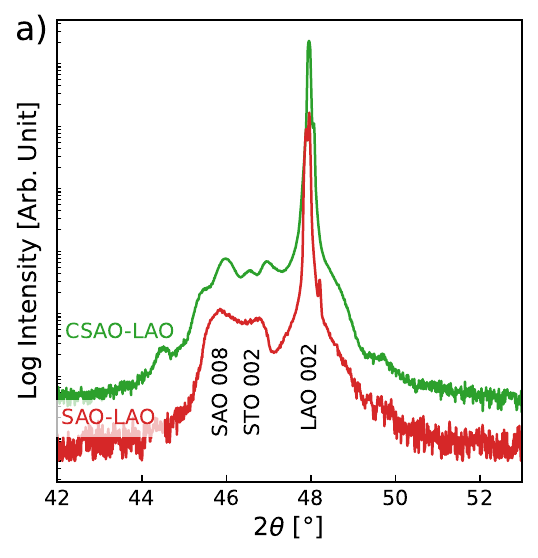} 
    \includegraphics[width=0.26\linewidth]{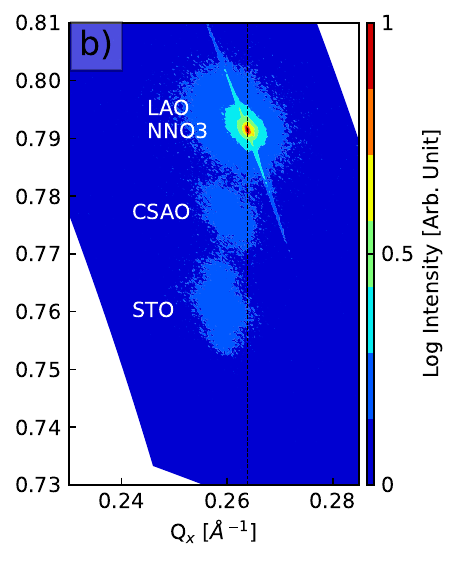}        
    \includegraphics[width=0.29\linewidth]{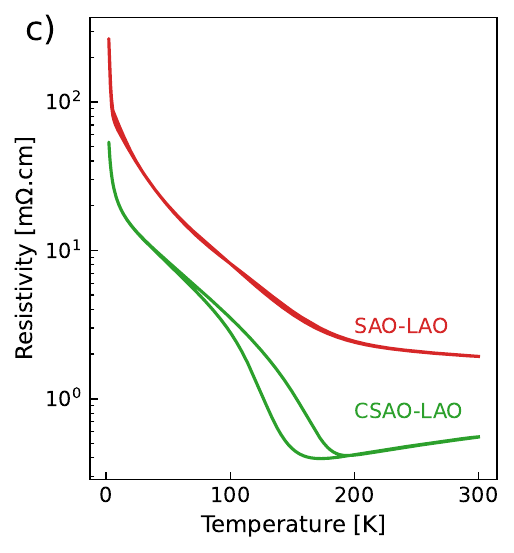}
    \caption{
    a) $2\theta-\theta$ scans of the as-grown CSAO-LAO and SAO-LAO heterostructures. 
    b) Reciprocal space map around the LAO -103 asymmetric reflections for the CSAO-LAO heterostructure.
    c) Transport measurements of the as-grown CSAO-LAO and SAO-LAO heterostructures.
}
    \label{fig:SI-th2th-rsm-epitaxial} 
\end{figure}
\begin{figure}[t!]
    \centering
    \includegraphics[width=0.28\linewidth]{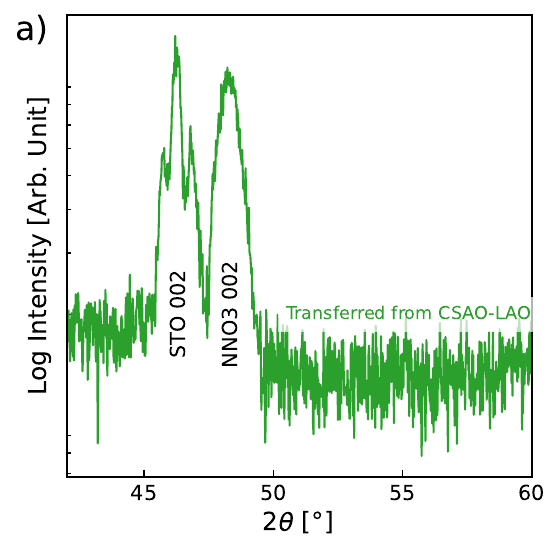}
    \includegraphics[width=0.25\linewidth]{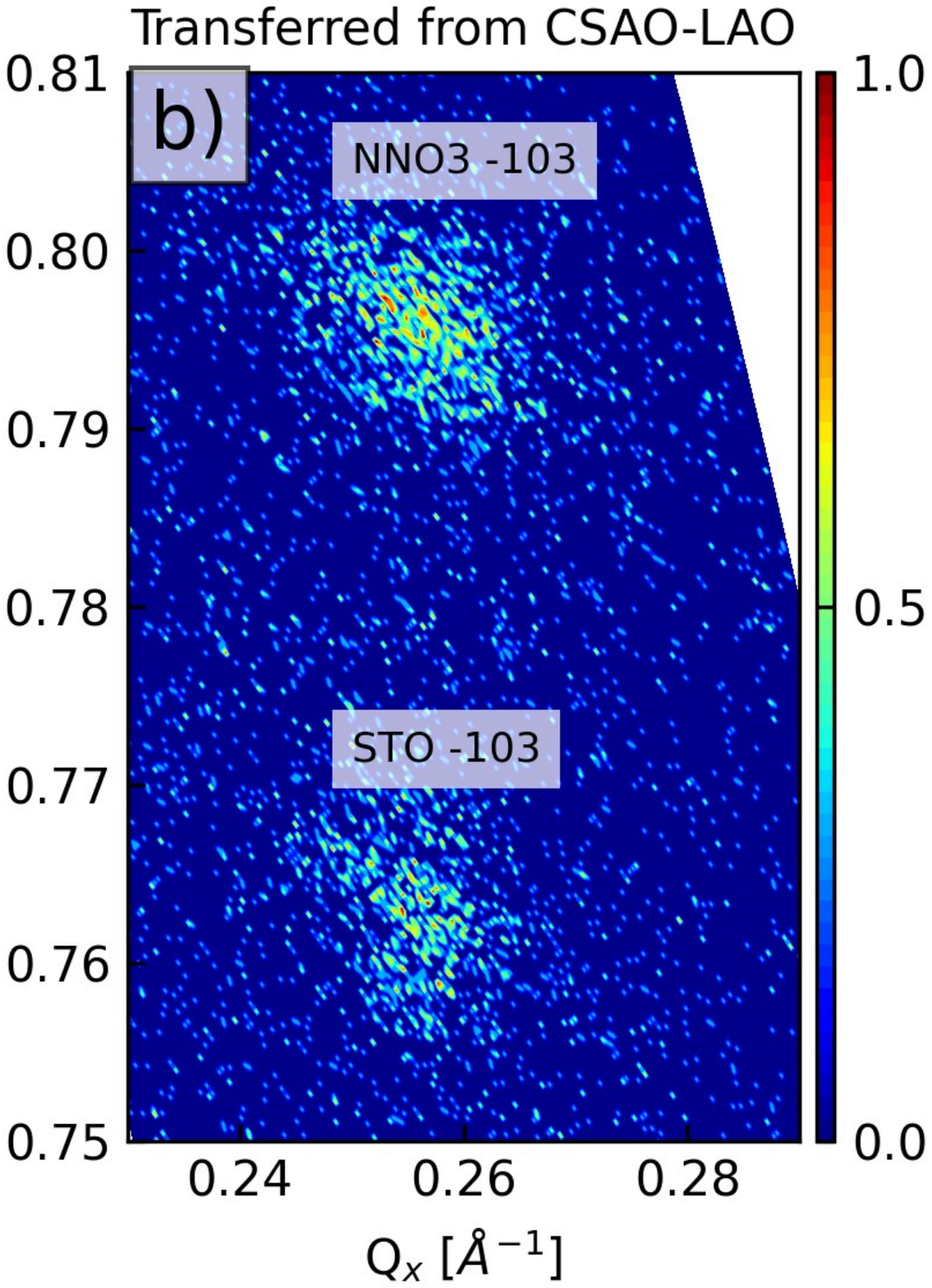}
    \includegraphics[width=0.38\linewidth]{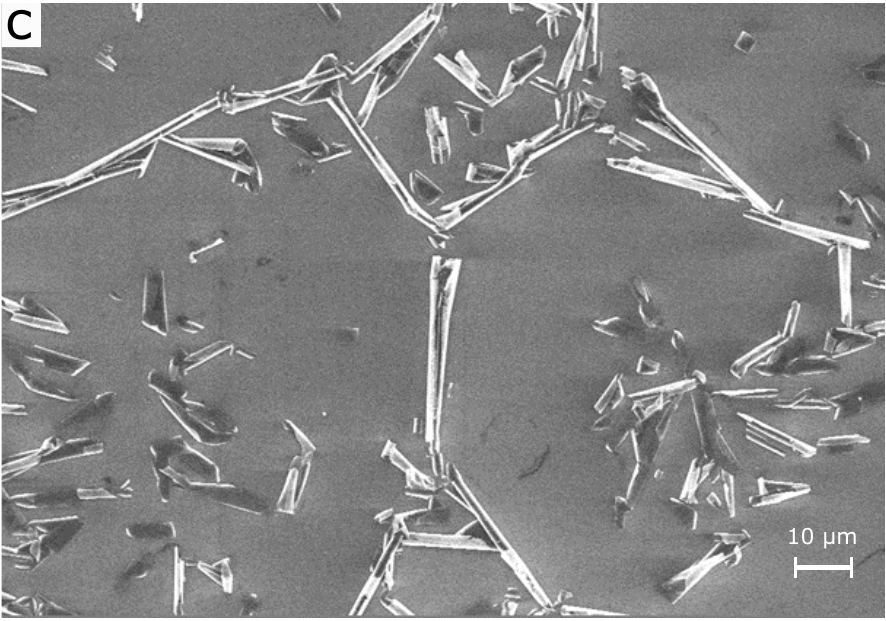}
    \caption{a) $2\theta-\theta$ scan,  b) Reciprocal space map around NNO3 and STO -103 reflection, and c) SEM image of the STO/NNO3/STO heterostructure trasferred from CSAO-LAO.}
    \label{fig:SI-th2thAndRsm-exfoliated}
\end{figure}
\begin{figure}[t!]
    \centering
    \includegraphics[width=0.35\linewidth]{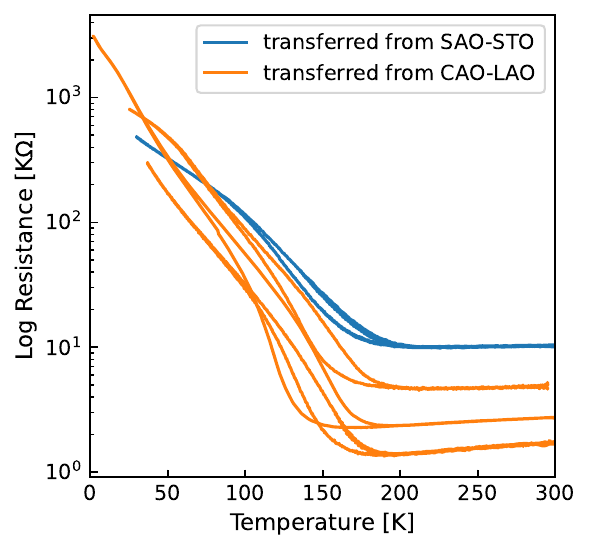}
    \caption{Transport measurements of the as-grown CAO-LAO and SAO-STO heterostructures for multiple membrane samples}
    \label{fig:SI-multiple-samples}
\end{figure}
\end{document}